\documentclass{elsart}
\usepackage{natbib}
\usepackage{graphicx}
\journal{New Astronomy Reviews}
\usepackage{amssymb}
\begin{document}
\begin{frontmatter}
\title{The shape of the CMB power spectrum.}
\author{Carolina J. \"Odman}
\ead{cjo25@mrao.cam.ac.uk}
\ead[url]{www.mrao.cam.ac.uk/\~{}cjo25/}
\address{MRAO, Cavendish Laboratory, Cambridge University, Cambridge, UK}
\begin{abstract}
The recent WMAP data represents a milestone in cosmology and helps
constrain cosmological parameters with unprecedented accuracy. In this work we
combine the WMAP data with previous CMB anisotropy measurements at smaller
angular scales to characterize the shape of the CMB anisotropy power spectrum.\\
We carry out a phenomenological analysis of the data. By allowing non-physical shapes
of the power spectrum we analyse high and low frequency experiments separately and together. 
We find that WMAP dramatically constrains the power spectrum up to
$\ell \sim 700$. On smaller scales, the data show discrepancies that can be associated with experimental systematics.
If we combine all types of experiments, the
observable features in the power spectrum are in excellent agreement with the
WMAP cosmological parameter estimation.\\
This work illustrates the advantages of a model-independent approach to
understanding experimental systematics that might affect CMB observations.
\end{abstract}
\begin{keyword}
Cosmic Microwave Background Anisotropies \sep Phenomenology \sep Observations
\PACS 98.80.Es
\end{keyword}
\end{frontmatter}
\vspace{-1cm}
\section{Introduction}\label{Intro}
\vspace{-0.3cm}
In recent years high resolution measurements of cosmic microwave background (CMB) anisotropies have measured the CMB power spectrum beyond the first acoustic peak. More recently the all-sky WMAP results represent a new step towards high precision cosmological parameter estimation. When combining these measurements with large-scale structure data and distant supernovae an increasingly consistent picture of the universe arises.

In order to investigate cosmological parameters in detail the experiments themselves have to be thoroughly understood. CMB anisotropy experiments are essentially of two kinds. High frequency experiments use bolometers and are usually balloon-born. The calibration, pointing and beam shape are the main sources of uncertainty of such instruments and tend to smear features in the power spectrum. Low frequency experiments use interferometers and direct;y measure the CMB in Fourier space. The calibration of such instruments is usually well understood and they do not suffer from beam uncertainty but the band-powers inferred from the observations tend to be anti-correlated. This can lead to increased contrast in the bands and is usually well encoded in the correlation matrix of the band-powers.

This work is an extension to a previous project presented in \cite{odman}. We use our previously obtained results as an input to a Markov Chain Monte-Carlo (MCMC) program to estimate the shape of the CMB anisotropy power spectrum using a phenomenological curve. We use the WMAP data along with other CMB data separated in three subsets according to the type of experiment. 
\vspace{-0.5cm}
\section{Method}\label{method}
\vspace{-0.3cm}
In recent cosmological parameter estimations based on CMB data (see \cite{melchpars} and references therein) it has been shown that the simplest theoretical models based on $6$ fundamental parameters are sufficient to describe the universe from currently available data. Common assumptions are a flat universe in which adiabatic initial perturbations evolve with no contribution from background gravity waves or massive neutrinos.

With the recently released WMAP data a more complicated cosmological picture can be tested \cite{spergel} including a varying spectral index, an equation of state of dark energy etc. It is has now become common to test a parameter space of $11$ dimensions or more. The shape of the CMB power spectrum allowed in such analyses is limited by physical models. It is therefore timely to investigate what model-independent shape is preferred by the data in order to test the influence of experimental systematics which might not be fully under control.
\begin{figure}[ht]
\begin{center}
\rotatebox{270}{\includegraphics[width=4.5cm]{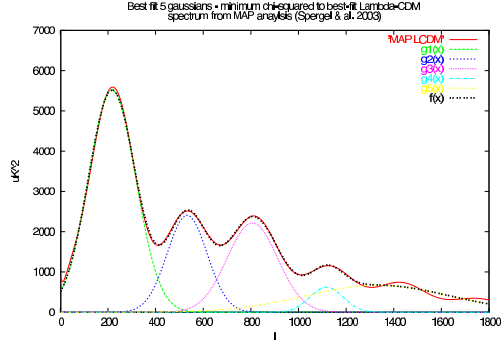}}
\end{center}
\caption{\label{gauss_fit}WMAP best-fit power spectrum fitted using a sum $f(x)$ of five gaussians $g_i(x)$. The fit is discrepant at very low multipoles because gaussians cannot reproduce the Sachs-Wolfe plateau and at high multipoles because there are not enough parameters. Between $50 \lesssim l \lesssim 1250$, the fit is excellent, showing the ability of our phenomenological curve to fit a physically calculated power spectrum.}
\end{figure}

In this analysis, we parametrise the CMB power spectrum using a sum of five consecutive Gaussians
\begin{equation}
\ell(\ell+1)C_\ell/2\pi = \sum_{i=1}^N \Delta T_i^2 \exp 
(-(\ell-\ell_i)^2/2\sigma_i^2), \ \  N = 5.
\end{equation}
\vspace{-0.3cm}
We then use an MCMC routine to constrain the fifteen parameters of our phenomenological curve; $\Delta T_i$, $l_i$ and $\sigma_i$. MCMC is an iterative process. After a certain burn-in time during which the Markov chains reach equilibrium, the parameters are sampled such that their number density is proportional to their posterior distribution. As explained in \cite{verde} such a routine is sensitive to the prior probability assumed for the parameters. Its efficiency also depends on the sampling strategy. If degeneracies exist among parameters it is useful to sample along degeneracy directions.

We make use of previously obtained results as an input to our MCMC routine for both the prior and the sampling strategy. We assume our previous results represent a $15$-dimensional Gaussian which we use a prior. For the sampling strategy, we diagonalise the covariance matrices obtained from our analysis before the WMAP data release. We then sample the parameter space as follows:
\begin{equation}
P_i = P_i^0 + \sum_j C_{ij} \lambda_j p_j, \ \  i, j = 1, ..., 15,
\end{equation}
\vspace{-0.7cm}
where $P_i$ are the parameters of our Gaussians, $\Delta T_i, l_i$ and $\sigma_i$. $P_i^0$ are the previously obtained best-fit values of those parameters, $\lambda_i$ are the eigenvalues of the covariance matrices, $C_{ij}$ is the matrix of eigenvectors and $p_j$ are sampled from a Gaussian of mean zero and unit variance. This is equivalent to following the axes of our Gaussian prior. We further check that the samples are within the previously assumed priors.

We then run our MCMC code for three subsets of data. We first separate high  (HF) and low frequency (LF) data as listed in table \ref{cmbdata} and then combine them. We always use the WMAP data. For each sample we calculate the phenomenological power spectrum and return its likelihood to the MCMC routine. The WMAP likelihood is calculated using the method described in \cite{verde}. For the other data the likelihood for a given phenomenological model is defined by  $-2{\rm ln} {\mathcal L}=(C_B^{ph}-C_B^{exp})M_{BB'}(C_{B'}^{ph}-C_{B'}^{exp})$ where  $M_{BB'}$ is the Gaussian curvature of the likelihood  matrix at the peak. When available we use the lognormal approximation to the band-powers. We marginalize over the reported Gaussian distributed  calibration error for each experiment and we include the beam uncertainties by the analytical marginalization method presented in \cite{sara}.
\begin{table}
\begin{center}
\caption{\label{cmbdata}List of experimental data used in this study. We use low and high frequency experiments both separately and together.\newline}
\begin{tabular}{lc|lc}
\hline\hline
High frequency (HF) data & $l$ range & Low frequency (LF)  data & $l$ range \\\hline
Acbar \cite{kuo}  & $150$ -- $3000$ & DASI  \cite{halverson} & $117$ -- $836$ \\ 
Archeops \cite{benoit}  & $15$ -- $350$ & CBI  \cite{pearson} & $400$ -- $1450$\\
Boomerang 98 \cite{ruhl}  & $50$ -- $1000$ & VSAE  \cite{grainge} & $160$ -- $1400$ \\ 
Maxima \cite{lee} & $73$ -- $1161$ \\\hline
\end{tabular}
\end{center}
\end{table}
\vspace{-0.5cm}
\section{Results and discussion}\label{results}
\vspace{-0.3cm}
We run our code until we obtain around $30,000$ samples for each run. The number density of each parameter is proportional to its marginalised posterior distribution. We summarize our resuls in table \ref{constraints} along with the pre-WMAP results for the fifteen Gaussian parameters of our phenomenological power spectra. The first peak is consistenly higher when including WMAP. It is shifted to slightly higher multipoles. The second peak is more pronounced than pre-WMAP when including all data. The third peak is very high for the LF data. This is mainly due to the VSAE band power at $\ell \sim 800$. At high multipoles, no distinct fourth or fifth peaks are detected, but the damping of the power is present in all cases.

\begin{table*}
\begin{center}
\caption{\label{constraints} 
1-$\sigma$ constraints on the phenomenological parameters before and after the inclusion of the WMAP results for three subsets of data.\newline}
{\scriptsize
\begin{tabular}{|c||c|c||c|c||c|c|}

\hline
Parameter & All data &  + WMAP & HF data &  + WMAP & LF data &  + WMAP\\\hline
$\Delta T_1 $& $69.3_{-2.3}^{+2.4} $&$72.19^{+0.13}_{-0.20}$& $70.7_{-3.5}^{+4.8} $ &$73.86^{+0.39}_{-0.34}$& $68.9_{-4.1}^{+5.5} $ &$73.91^{+0.25}_{-0.28}$\\
$\Delta T_2 $& $44.0_{-1.9}^{+1.7} $&$49.77^{+1.28}_{-0.47}$& $43.1_{-3.0}^{+3.6} $ &$42.73^{+2.02}_{-0.17}$& $46.8_{-4.3}^{+3.6} $ &$48.52^{+0.41}_{-0.61}$\\
$\Delta T_3 $& $43.8_{-4.4}^{+2.1} $&$42.78^{+0.40}_{-0.07}$& $40.7_{-5.0}^{+4.3} $ &$41.09^{+0.09}_{-0.63}$& $54.4_{-11.2}^{+3.4} $ &$54.47^{+0.30}_{-0.44}$\\
$\Delta T_4 $& $31.2_{-8.7}^{+5.7} $&$38.58^{+0.18}_{-1.08}$& $25.7_{-5.7}^{+3.1} $ &$31.59^{+0.00}_{-0.25}$& $27.9_{-7.3}^{+8.7} $ &$38.69^{+0.11}_{-0.56}$\\
$\Delta T_5 $&  $19.8_{-3.8}^{+1.8} $ &$16.14^{+0.31}_{-1.04}$& $18.4_{-5.2}^{+3.2} $ &$20.74^{+1.46}_{-0.19}$& $20.6_{-17.3}^{+12.7} $ &$23.66^{+0.36}_{-1.15}$\\
$\ell_1$&$208.8_{-6.1}^{+6.2}$&$214.3^{+0.8}_{-0.7}$& $204.6_{-7.9}^{+11.4}$ &$215.8^{+0.2}_{-0.6}$& $206.8_{-22.0}^{+10.8}$ &$216.7^{+0.3}_{-0.9}$\\
$\ell_2$&$550_{-45}^{+13}$&$548.5^{+0.1}_{-1.6}$& $505_{-21}^{+25}$ &$512.4^{+0.1}_{-0.1}$& $533_{-20}^{+25}$ &$533.2^{+1.4}_{-1.4}$\\
$\ell_3$&$824_{-41}^{+12}$&$831.5^{+5.0}_{-1.3}$& $764_{-42}^{+74}$ &$744.5^{+0.1}_{-1.2}$& $806_{-36}^{+26}$ &$803.0^{+1.1}_{-8.3}$\\
$\ell_4$& $1145_{-45}^{+30}$ &$1219^{+4}_{-6}$& $1158_{-67}^{+242}$ &$1241^{+2}_{-3}$& $1189_{-87}^{+32}$ &$1195^{+2}_{-3}$\\
$\ell_5$& $1474_{-79}^{+153}$ &$1501^{+0}_{-6}$& $1649_{-262}^{+142}$ &$1735^{+2}_{-1}$& $1515_{-346}^{+81}$ &$1577^{+2}_{-1}$\\
$\sigma_1$&$93.3_{-5.2}^{+4.5}$&$99.0^{+0.4}_{-0.7}$& $90.3_{-6.2}^{+8.2}$ &$99.8^{+0.7}_{-1.0}$& $88.2_{-12.3}^{+12.7}$ &$99.3^{+0.5}_{-0.8}$\\
$\sigma_2$&$111.2_{\mbox{\tiny n. c.}}^{+27.7} $&$103.1^{+0.5}_{-7.1}$& $78.2_{-12.2}^{+19.3}$&$61.4^{+3.6}_{-1.0}$& $61.9_{\mbox{\tiny n. c.}}^{+36.7}$ &$82.8^{+0.2}_{-1.2}$\\
$\sigma_3$&$82.5_{-23.1}^{+20.7}$&$< 58.8$& $136.3_{\mbox{\tiny n. c.}}^{+94.2}$ &$171.4^{+1.2}_{-0.7}$& $69.8_{\mbox{\tiny n. c.}}^{+12.8}$ &$<56.8$\\
$\sigma_4$& n. c. &$> 248.5$& n. c. &$248.6^{+0.4}_{-1.3}$& n. c. &$198.5^{+1.5}_{-1.5}$\\
$\sigma_5$& n. c. &$182.5^{+0.3}_{-11.0}$& n. c. &$198.9^{+0.9}_{-0.6}$& n. c. &$60.5^{+6.6}_{-4.7}$\\
\hline
\end{tabular}}
\end{center}
\end{table*}

\begin{figure}[ht]
\begin{center}
\includegraphics[width=4.5cm]{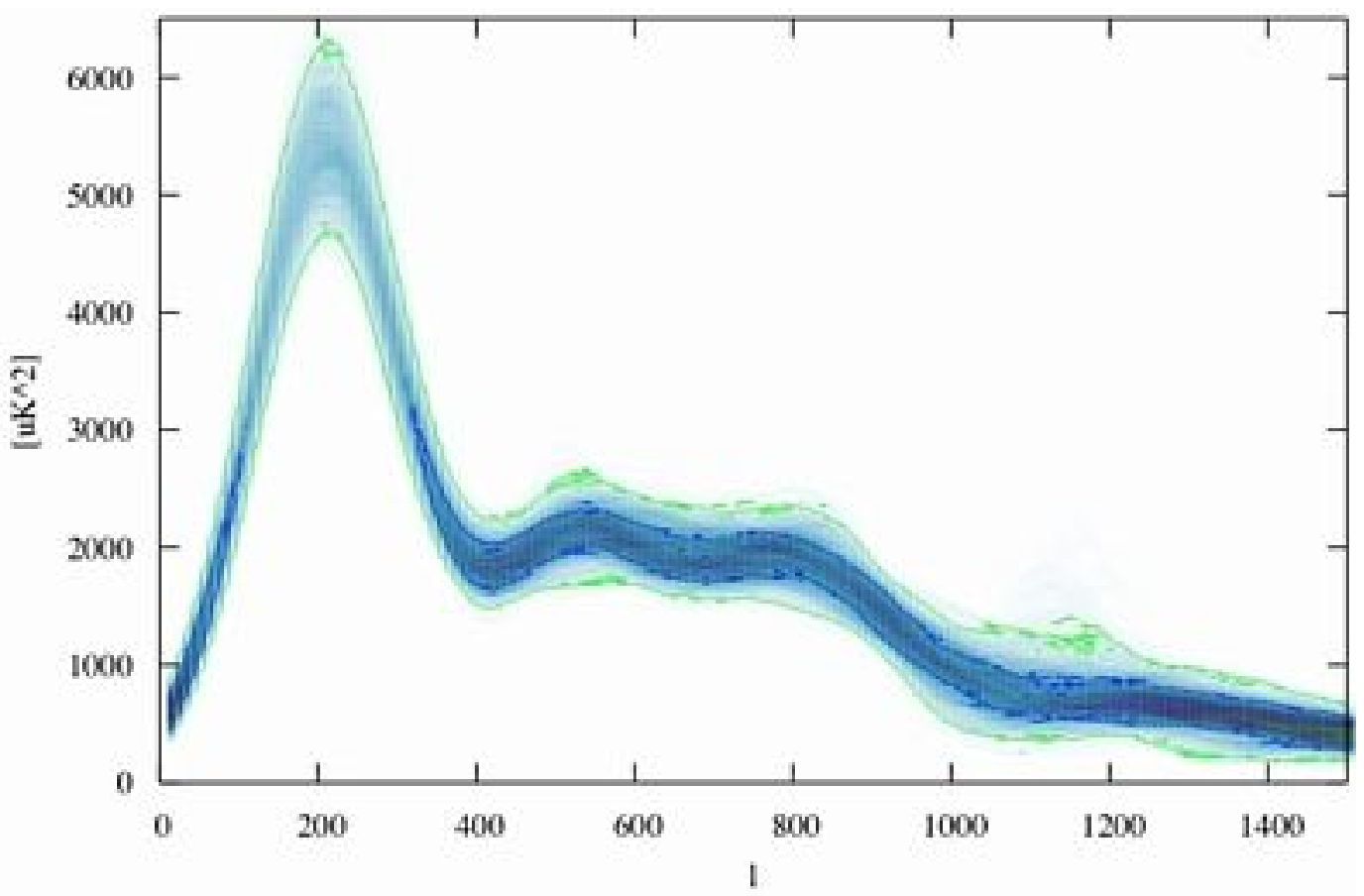}\ \includegraphics[width=4.5cm]{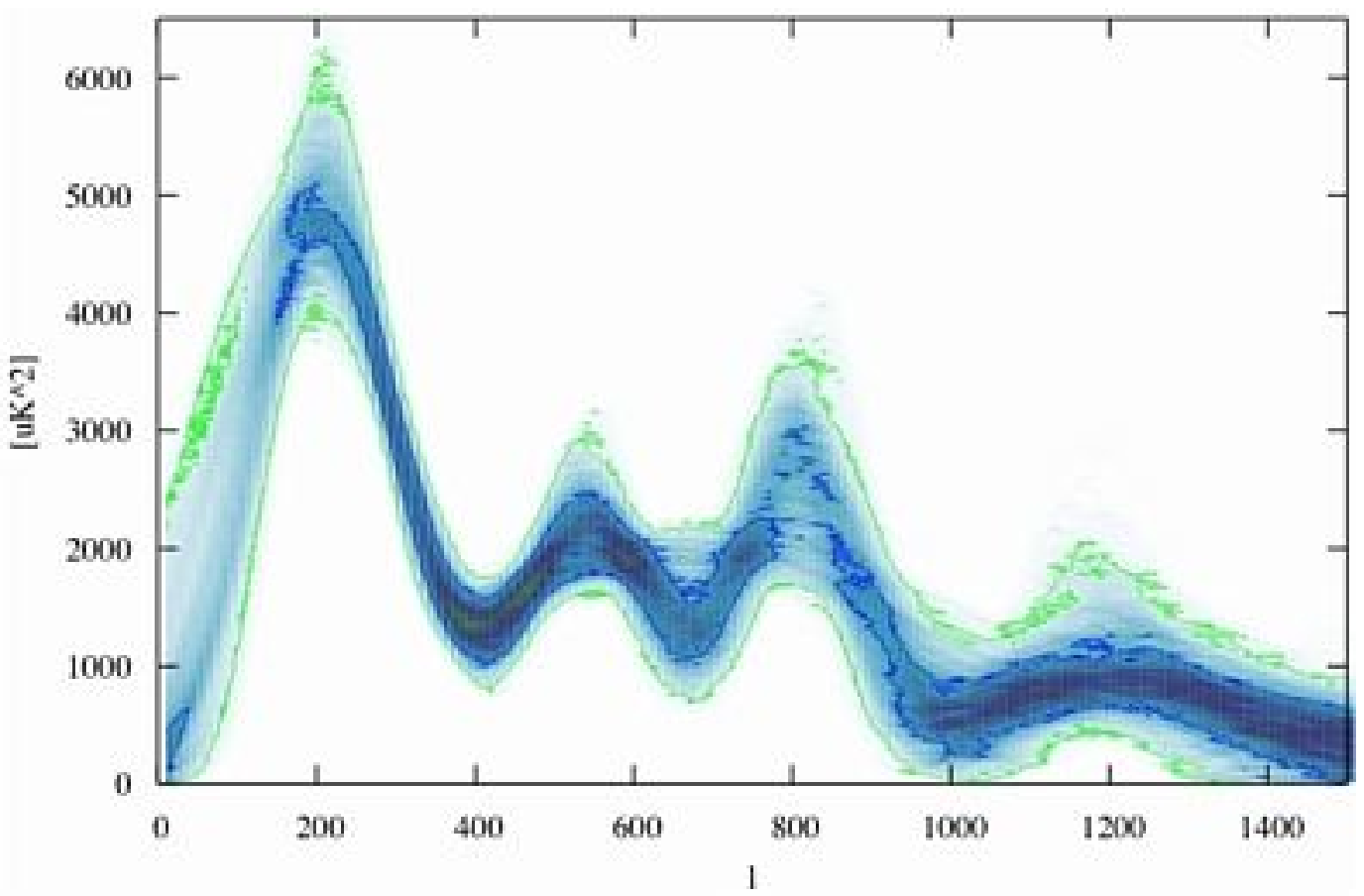}\ \includegraphics[width=4.5cm]{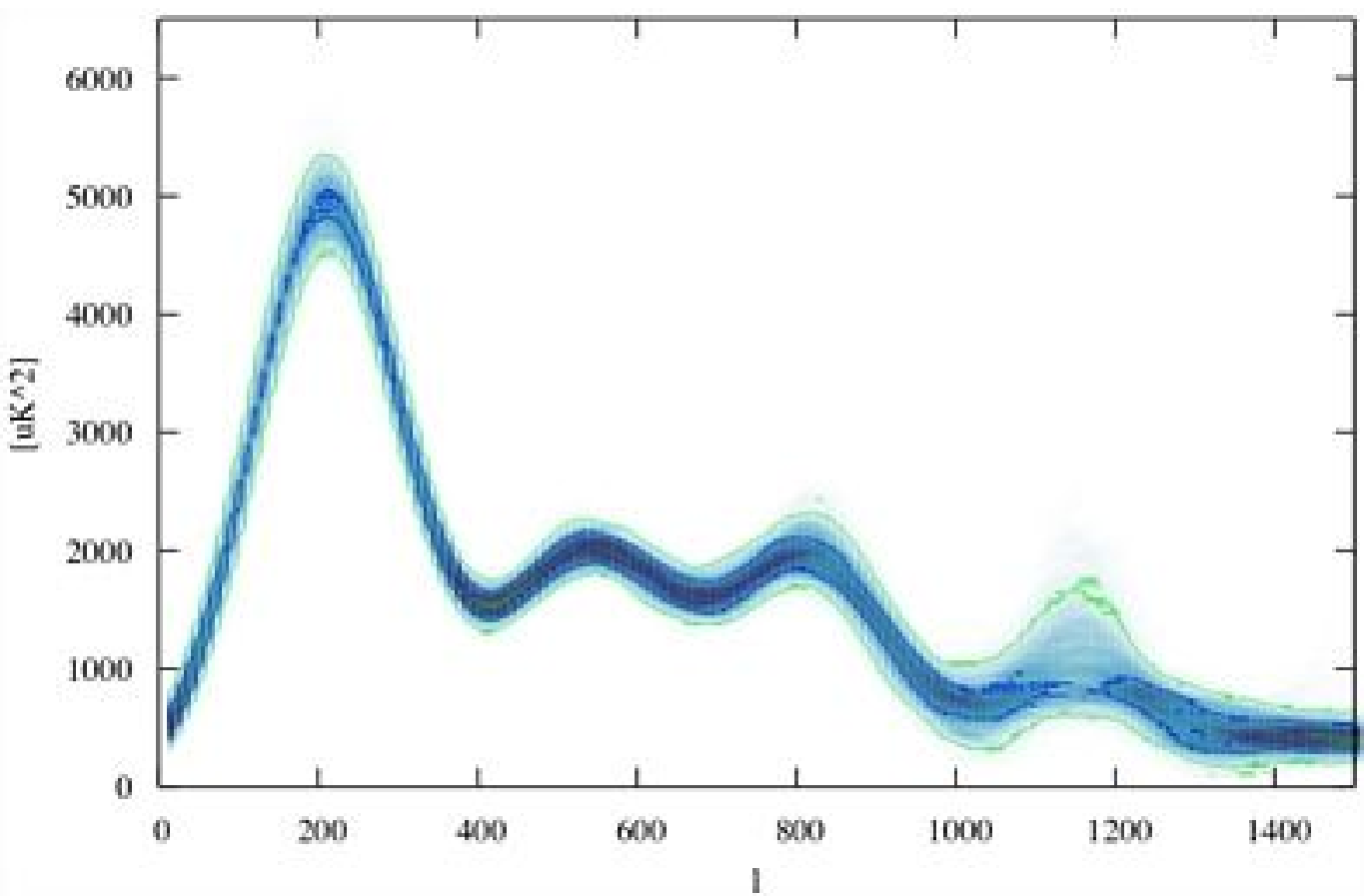}\newline
\includegraphics[width=4.5cm]{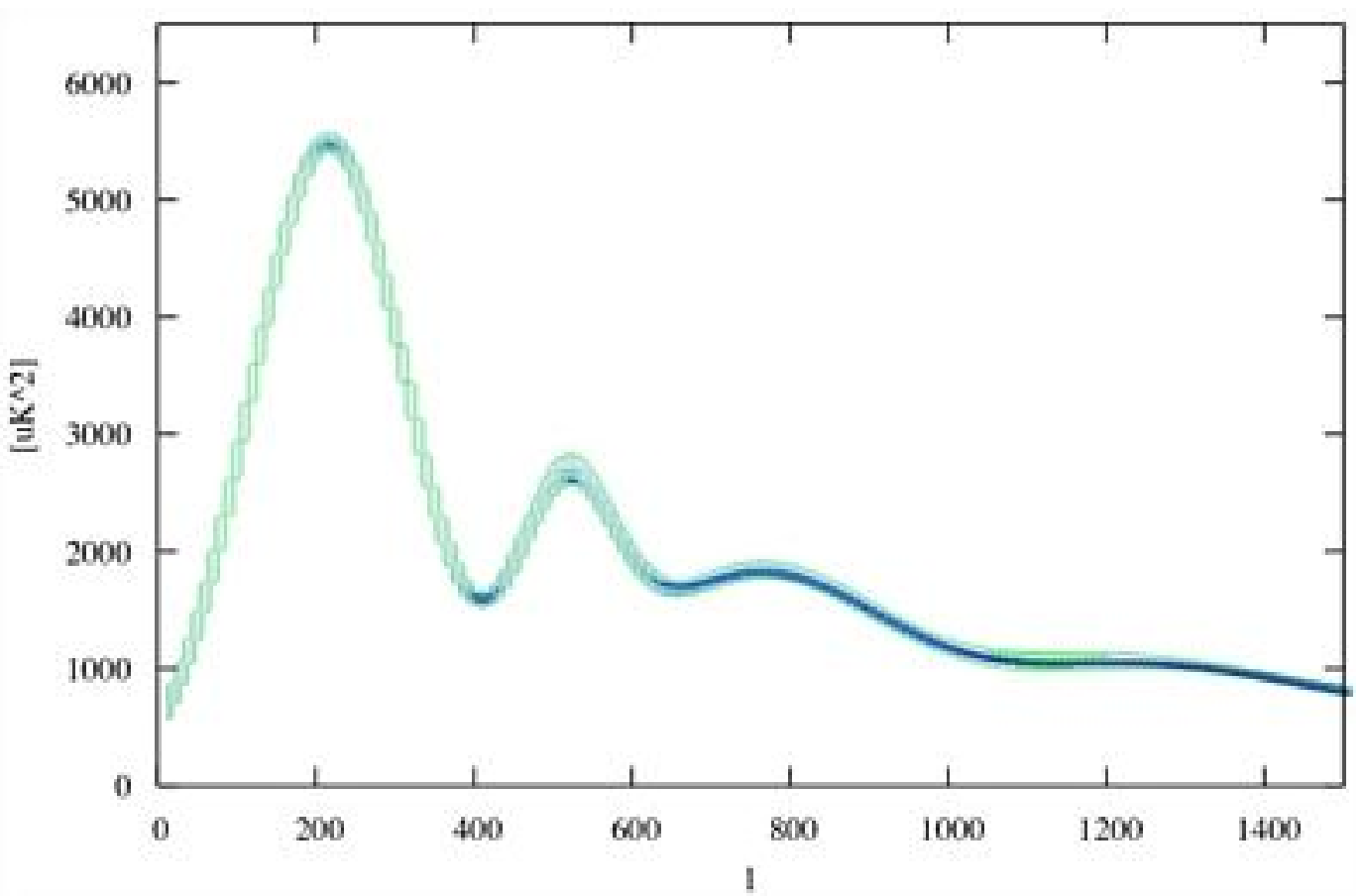}\ \includegraphics[width=4.5cm]{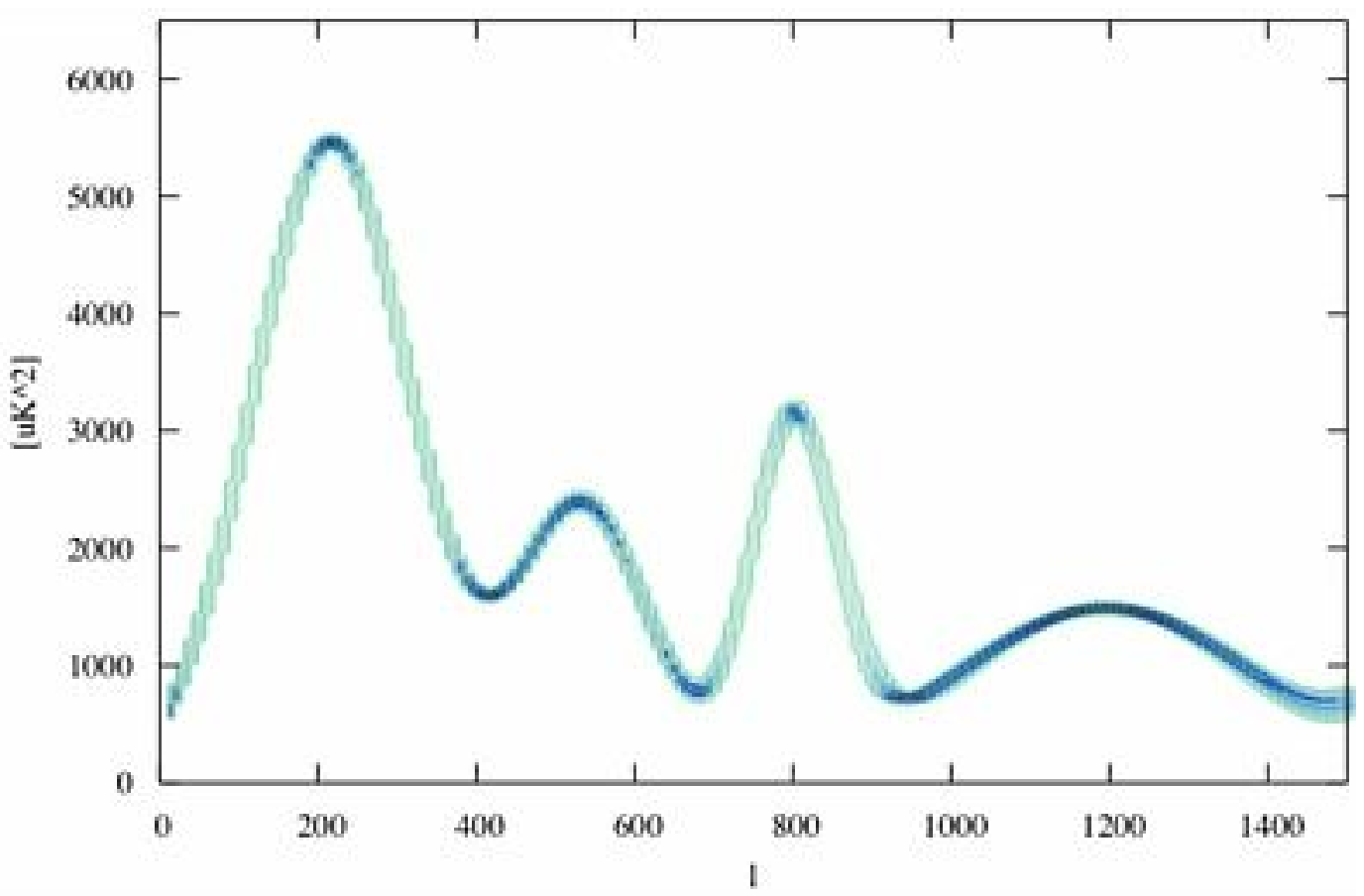}\ \includegraphics[width=4.5cm]{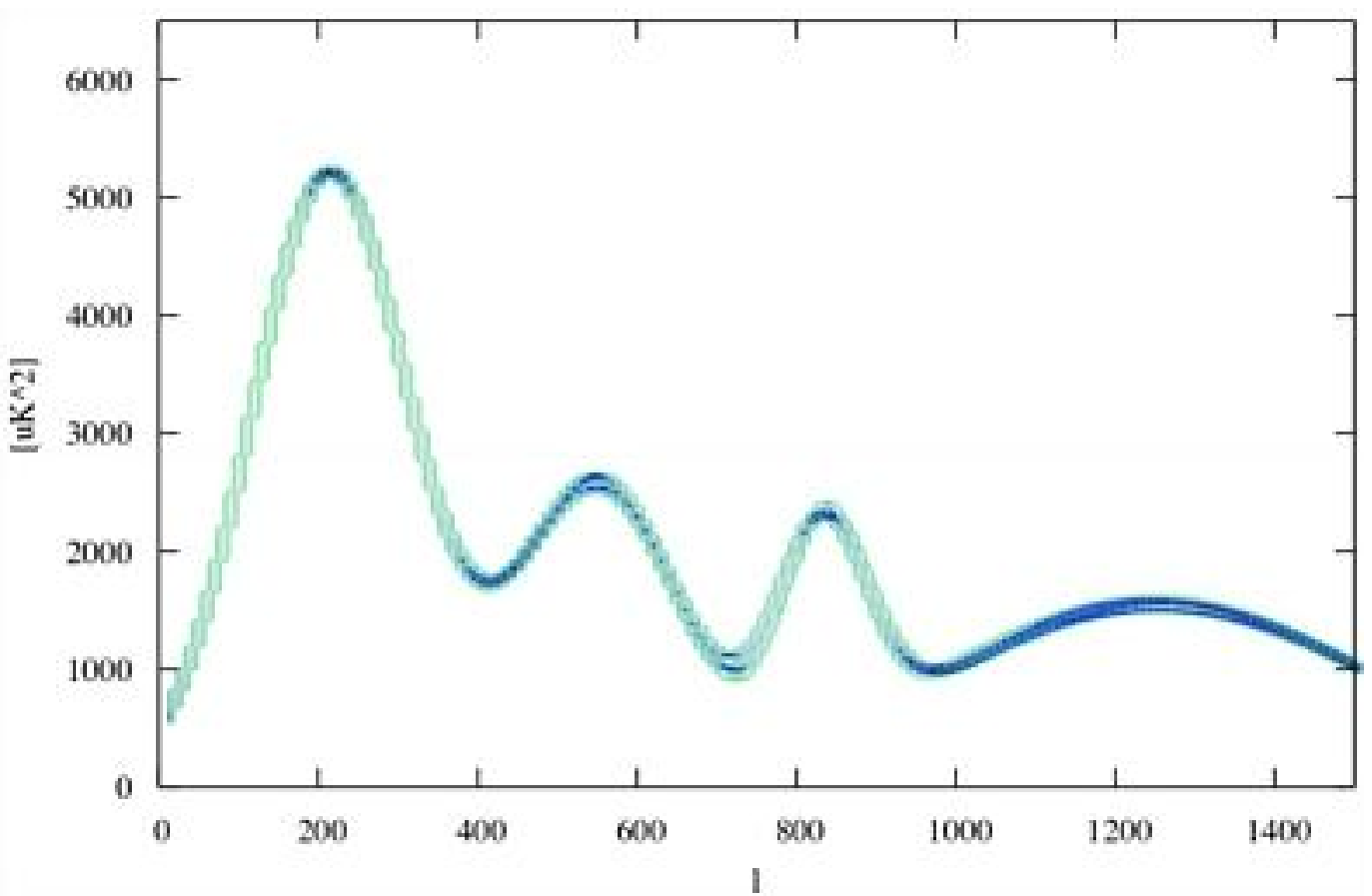}
\end{center}
\caption{\label{tsunami}Shape of the phenomenological CMB power spectrum preferred by the three subsets of data. The first row corresponds to pre-WMAP results which were used as a prior for the analysis including WMAP data which is shown on the second row. The first column is HF data, the second column is LF data and the third column is all data. The green contours are 2-$\sigma$ confidence levels.}
\end{figure}

In order to get the shape of the power spectrum we divide the $\ell$ - $\mu K^2$ space into pixels of size $\Delta l = 10$, $\Delta T = 10$ and count the number of phenomenological spectra that go through each pixel. The resulting allowed shapes are shown in figure \ref{tsunami} along with pre-WMAP results. This confirms the consistency and accuracy of the constraints obtained on the first two peaks. At smaller angular scales, as we had obtained before, the discrepancy remains between high  and low frequency experiments. Even if we include correlations and marginalise over calibration and beam uncertainties in the data the high frequency data show less contrasted features in the power spectrum than low frequency experiments.

\begin{figure}[ht]
\begin{center}
\includegraphics[width=5cm]{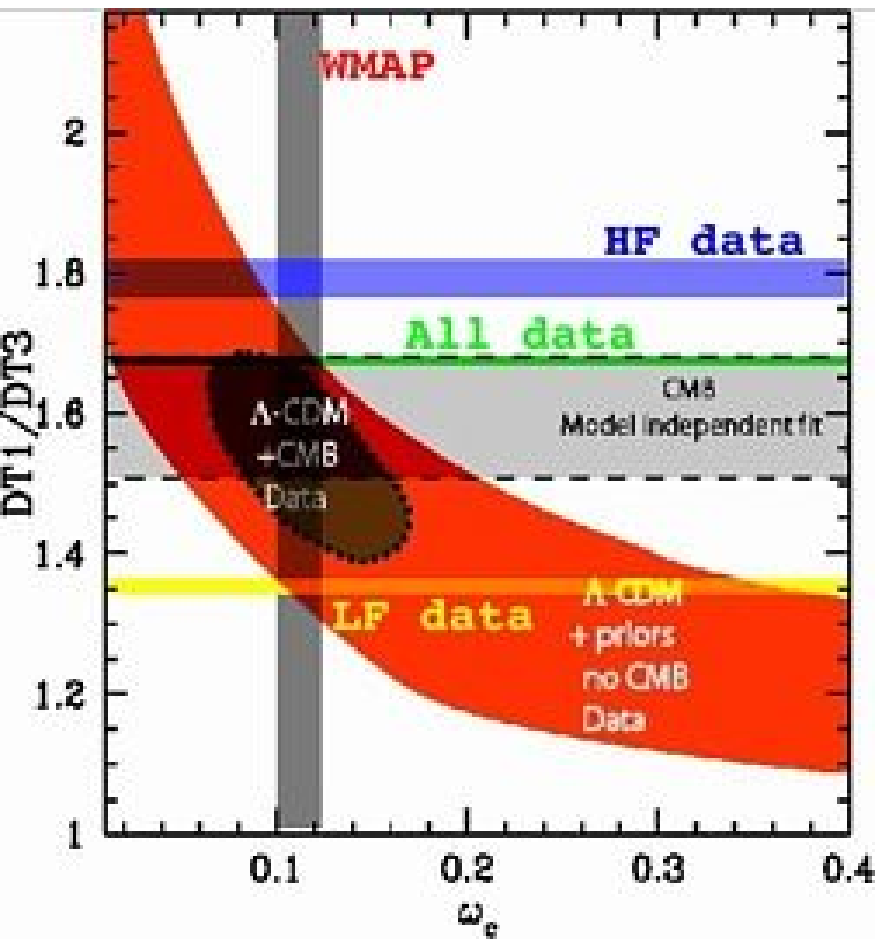}\ \ \ \ \includegraphics[width=5cm]{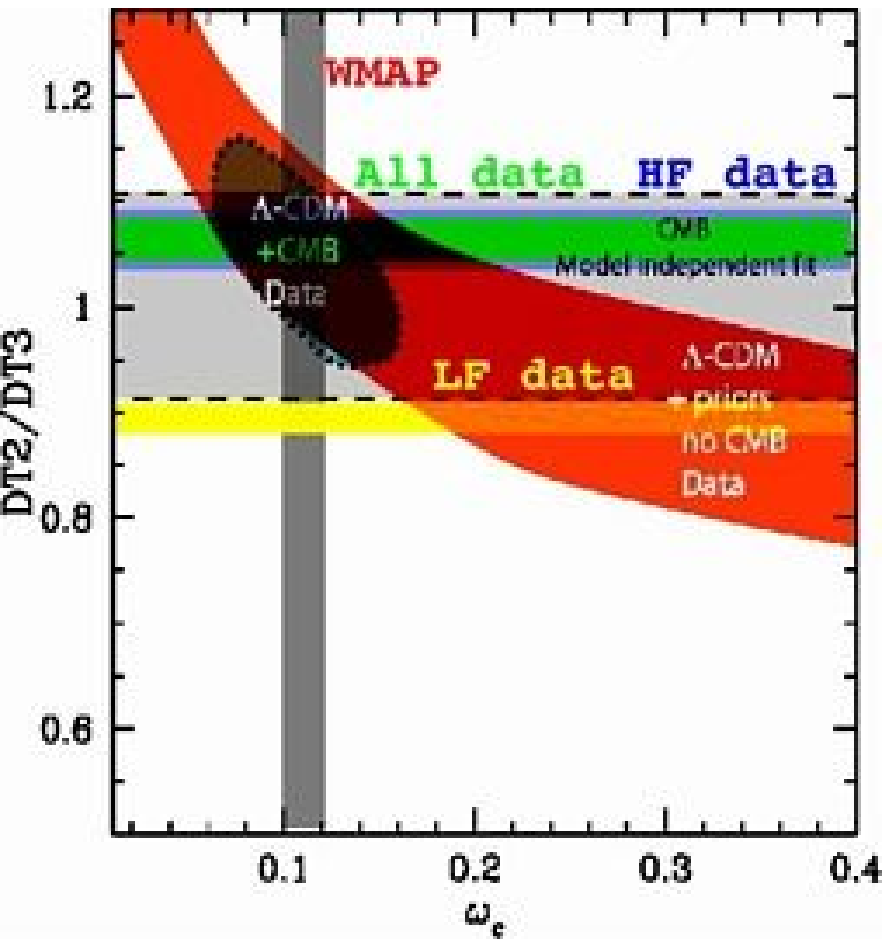}
\end{center}
\caption{\label{cosmopars}Ratio of amplitudes of the third peak to the first and second peaks vs the physical density of cold dark matter. Our phenomenological results differ. However, when combining all data, we are consistent with the WMAP estimation (vertical gray region). The orange region corresponds to the degeneracy direction for physical models. The grey horizontal region corresponds to pre-WMAP results and the brown region corresponds to a pre-WMAP parameter estimation. Our results for the HF, LF and all data are shown in the blue, yellow and green horizontal strips, respectively.}
\end{figure}

In particular the third peak is higher than the second peak for low frequency experiments. This has important consequences in terms of the estimate of cold dark matter in the universe as shown in figure \ref{cosmopars}. The yellow region in figure \ref{cosmopars} which corresponds to the low frequency experiments intersects the orange regions for higher values of the CDM physical density than from the WMAP parameter estimation (vertical gray region). The ratio of the first to the third peak from high-frequency data allows for a lower CDM density owing to the less marked secondary features in the power spectrum.

\vspace{-0.5cm}
\section{Conclusions}\label{conclusions}
\vspace{-0.3cm}
In this work, we take advantage of the power of the WMAP data to constrain the shape of the CMB power spectrum. We find that when including data from both high and low frequency experiments the preferred shape of the power spectrum is very close to the shape of the physical `concordance model'. Low frequency data show very contrasted second and third peaks. High frequency data prefer more shallow features. In all cases a damping of the power on small scales is detected.

Such a model-independent approach sheds light on discrepancies between high and low frequency experiments which would be less marked in traditional parameter estimation techniques. Our best fit shapes are not physical but the main observables of the spectrum are consistent with the WMAP cosmological parameter estimations.

Further investigation of the experiment-related discrepancies would be useful and WMAP provides a strong starting point. In combination with higher multipole data the shape of the CMB power spectrum is now strongly constrained. There is no doubt that a second and third peaks have been detected. This illustrates the new era cosmology has entered: the question now is not whether secondary peaks are present, thereby strengthening the already compelling agreement with inflationary predictions but how strong those secondary features are.

We would like to thank Alessandro Melchiorri, Anthony Lasenby, Licia Verde, Hiranya Peiris and Rob Izzard for useful comments. CJ\"O is supported by an Isaac Newton Studentship and a Girton College Scholarship.

\end{document}